\documentstyle[prd,preprint,eqsecnum,aps,tighten]{revtex}
\begin{document}
\title{Mode-by-mode summation for the zero point
electromagnetic  energy  of an infinite cylinder}
\draft
\preprint{OKHEP-98-09}
\author{Kimball A.~Milton\thanks{Electronic address:
milton@mail.nhn.ou.edu}}
\address{Department of Physics and Astronomy, University of Oklahoma,
Norman, Oklahoma 73019, USA }
\author{A.~V.~Nesterenko}
\address{Physical Department, Moscow State University, Moscow,
119 899, Russia}
\author{ V.~V.~Nesterenko\thanks{Electronic address:
nestr@thsun1.jinr.dubna.su}}
\address{Bogoliubov Laboratory of Theoretical Physics, Joint
Institute for
Nuclear Research \\ Dubna, 141980, Russia}
\date{\today}
\maketitle
\begin{abstract}
     Using the mode-by-mode summation technique the zero point
energy of the electromagnetic field is calculated for the boundary
conditions  given on the surface of an infinite solid
cylinder.  It is assumed that the  dielectric and magnetic
characteristics of the material which makes up the cylinder
$(\varepsilon_1, \,\mu_1)$ and of that which makes up
the surroundings $(\varepsilon _2,\,\mu_2)$ obey the relation
$\varepsilon _1\mu_1= \varepsilon _2\mu_2$. With this assumption all
the divergences cancel. The divergences are regulated by making use of zeta
function techniques. Numerical calculations are carried out for a
dilute dielectric-diamagnetic cylinder and for a perfectly conducting
cylindrical shell. The Casimir energy in the first case vanishes,
and in the second is
in a complete agreement  with that obtained by DeRaad and Milton
who employed a Green's function technique with an ultraviolet regulator.
\end{abstract}
\section{INTRODUCTION}
     Calculation of the Casimir energy for nontrivial boundary
conditions is a subject of intense ongoing activity. In spite of
this, the experience accumulated in this area still does not allow
one  to predict, without involved calculation, even the sign of this
energy~\cite{BD,Milton}. In this connection the development of new
effective methods of  calculating  the Casimir energy is doubtless of
interest.

     In recent papers~\cite{Sphere,Ball} the zero point energy of the
electromagnetic field with boundary conditions given on a sphere has
been calculated by a method of direct summation of the eigenfrequencies.
A substantial point here was the use of contour integration in
the complex frequency plane~\cite{zeta}. The divergences in this
problem were removed by two subsequent steps. The first was accomplished by
the subtraction of the vacuum energy of an infinite homogeneous space
and the second made use of the technique of zeta function regularization.
The same method was also applied for calculating the
Casimir energy of a massless scalar field obeying Dirichlet and
Neumann boundary conditions on a sphere~\cite{Sphere}.
The same technique has recently been used to rederive the electromagnetic
Casimir energy for a spherical shell~\cite{hagen}.
The possibility of incorporating in this approach the dielectric and
magnetic properties of the media was demonstrated in~\cite{Ball}.
These calculations are distinguished by a certain concision and simplicity in
comparison with the Green's function techniques~\cite{Milton}, which,
however, retain superiority with respect to physical interpretation.

     The present paper seeks to show the efficiency of the mode
summation method in calculating the zero point energy of
an electromagnetic field when the boundary conditions are given on an
infinite circular cylinder. When applying the Green's function
techniques~\cite{Milton,BN,Raad}, this problem turns out to be more
complicated than the corresponding calculation of the Casimir energy for
sphere~\cite{MRS,Miltonb,ML,MiltonNg,MiltonNg2,BrevikK,Brevikcor}.

     The layout of the paper is as follows. In Sec.~II the general
integral representation is derived for the Casimir energy of an
infinite solid cylinder surrounded by an uniform medium.  The
permittivity and permeability of the cylinder material
$(\varepsilon_1,\,\mu_1)$ and those  of the surroundings
$(\varepsilon_2,\,\mu_2)$ are considered to be arbitrary. In
principle they may depend on the frequency of the electromagnetic
oscillations (dispersion of the media), but this point is beyond the
scope of the present paper. In Sec.~III it is assumed that the
electromagnetic characteristics obey the condition
$\varepsilon_1\mu_1=\varepsilon_2\mu_2=c^{-2}$ where $c$ is the speed of
the light in the media (in units of that in the vacuum). When this condition is
satisfied, all the divergences cancel between interior and exterior modes.
Those divergences are regulated by employing the zeta
function technique. In Sec.~IV the cases
when $\xi^2 \ll 1$ and $\xi ^2=1$ are considered numerically, $\xi ^2$ being
$(\varepsilon _1-\varepsilon _2) ^2/(\varepsilon _1+\varepsilon
_2)^2$. The first case gives the Casimir energy of a dilute
dielectric-diamagnetic cylinder, while the second case corresponds to  a
perfectly conducting infinitely thin cylindrical shell. 
Remarkably, the Casimir energy obtained for a tenuous medium vanishes,
as it does for a tenuous dielectric cylinder.
The result obtained in the second case
is identical to that obtained by the Green's function method
of calculating the energy and regulating the divergences 
by use of an ultraviolet regulator \cite{Milton}.  
In Sec.~V (Conclusion) the significance of the universal results obtained
are discussed.

\section{Integral representation for the Casimir
energy}
     We shall consider the following configuration. An infinite
cylinder of radius $a$ is placed in an uniform unbounded
medium. The permittivity and the permeability of the material making up
the cylinder are $\varepsilon _1$ and $\mu _1$, respectively, and those for
surrounding medium are $\varepsilon _2$ and $\mu _2$. It is assumed that
the conductivity in both the media is zero. We will compute the
Casimir energy per unit length of the cylinder.

In the mode summation method, the renormalized Casimir energy is
defined by
\begin{equation}
\label{definition}
E=\frac{1}{2}\sum_{\{p\}}(\omega _p-\bar \omega _p),
\end{equation}
where $\omega _p$ are  the classical eigenfrequencies of the
electromagnetic oscillations in the system under consideration, and
$\bar \omega_p$ are those in the absence of any boundary, that is,
when either medium fills all space. (When the precise meaning is not
required, we denote this by the formal limit
 $a \to \infty$.)  The set $\{p\}$ stands for  a
complete set of quantum numbers (discrete and continuous) which is
determined by the  symmetry of the problem.  Either sum in Eq.\
(\ref{definition}) diverges, therefore a preliminary regularization
is required.

      In order for the eigenfrequencies to be found one needs to
solve Maxwell's equations for the given configuration with allowance
for the appropriate boundary conditions on the lateral surface of the
cylinder. As is well known, it is sufficient to  require the continuity of
the tangential components of the electric field ${\bf E}$ and
of the magnetic field ${\bf H}$ \cite{Stratton}.
In terms of the cylindrical coordinates
$(r,\theta,z)$ the eigenfunctions of the given boundary value problem
contain the multiplier
\begin{equation}
\exp(-i\omega t+ik_z z+in\theta),
\end{equation}
and their dependence on $r$ is described by the cylindrical Bessel
functions $J_n$ for $r<a$ and by the Hankel functions of the first
kind $H_n\equiv H_n^{(1)}$ for $r>a$. The eigenfrequencies are the
roots of the equation (Ref.~\cite{Stratton}, p.~526)
\begin{eqnarray}
\label{general} f_n(k_z,\omega,a)=0,
\end{eqnarray}
where
\begin{eqnarray}
f_n\equiv \lambda _1^2\lambda_2^2 \Delta_n^{\text{TE}}(\lambda_1a,
\lambda_2a)\, \Delta_n^{\text{TM}}(\lambda_1a, \lambda_2a)
-n^2\omega^2k^2_z(\varepsilon_1\mu_1
-\varepsilon_2\mu_2)^2\left(J_n(\lambda_1a)\,H_n(\lambda_2a)
\right )^2,
\label{general1}
\end{eqnarray}
with
\begin{eqnarray}
\Delta_n^{\text{TE}}(\lambda_1a, \lambda_2a)&=&
a \mu_1 \lambda_2 J_n'(\lambda_1 a)\,H_n(\lambda_2a)-
a \mu_2 \lambda_1 J_n(\lambda_1 a)\,H'_n(\lambda_2a), \nonumber \\
\Delta_n^{\text{TM}}(\lambda_1a, \lambda_2a)&=&
a \varepsilon_1 \lambda_2 J_n'(\lambda_1 a)\,H_n(\lambda_2a)-
a \varepsilon _2 \lambda_1 J_n(\lambda_1 a)\,H'_n(\lambda_2a),
\end{eqnarray}
\[
\lambda_i^2=k_i^2-k_z^2,\quad k_i^2=\varepsilon _i\mu_i \omega^2,
\quad i=1,2, \quad n=0,\pm 1,\pm 2, \ldots .
\]
The indices TE and TM will be explained below. The prime on the
functions $J_n$ and $H_n$ means differentiation with respect to their
arguments. For given $k_z$ and $n$ Eq.~(\ref{general}) has an
infinite sequence of roots $\omega_{nm}(k_z),\;m=1,2,\ldots $,
these frequencies being the same inside and outside the
cylinder~\cite{endnote}. In view of this the Casimir energy
(\ref{definition}) can be rewritten as
\begin{equation}
\label{sum}
E=\frac{1}{2}\int_{-\infty}^{\infty}\frac{dk_z}{2\pi}\sum_{n=
-\infty}^{\infty}
\sum_{m=1}^{\infty}\left [\omega_{nm}(k_z)-\bar\omega_{nm}(k_z)
\right ],
\end{equation}
where $\bar \omega_{nm}(k_z)$ stands for the uniform medium subtraction
referred to above.

The next step in our consideration is a representation of the sum
in Eq.~(\ref{sum}) in terms of the contour integral~\cite{zeta}
\begin{equation}
\label{contour}
E=\frac{1}{2}\int_{-\infty}^{\infty}\frac{dk_z}{2\pi}\sum_{n=
-\infty}^{\infty}\frac{1}{2\pi i}\frac{1}{2}
\oint _C\omega \,d_\omega \ln \frac{f_n(k_z,\omega,a)}
{f_n(k_z,\omega,\infty)}.
\end{equation}
 Integration in (\ref{contour}) is conducted
along a closed path $C$ in the complex $\omega$ plane which  
consists of two parts: $C_+$ which encloses
the positive roots of Eq.~(\ref{general}) in a counterclockwise sense,
and $C_-$ which encircles the negative roots in a clockwise sense.
Therefore, we face to task of investigating the properties of the function
$f_n(k_z,\omega ,a)$ specifying the frequency equation.
(However, the result (\ref{contour}) may be easily shown to be
equivalent to the corresponding Green's function formulation.
See Appendix A.) Generally
this is a problem of extreme difficulty. Therefore, in the next
sections we shall consider specific cases introducing simplifying
assumptions.

The method of calculation of the Casimir energy proposed above can
straightforward be generalized  to the dispersive media. To this end,
it is sufficient to treat the parameters $\varepsilon _i$ and $\mu_i,
\; i=1,2$ in the frequency equation (\ref{general}) as given functions of
the frequency~$\omega$. However, this issue is beyond the scope of the
present paper.

\section{Casimir energy of an infinite cylinder when
$\varepsilon_1\mu_1=\varepsilon_2\mu_2$}
We assume that the permittivity and permeability of the cylinder material
$(\varepsilon_,\, \mu_1)$ and of the surroundings
$(\varepsilon_2,\,\mu_2)$ are not arbitrary but  satisfy the
condition
\begin{equation}
\label{condition}
\varepsilon_1\mu_1=\varepsilon_2\mu_2=c^{-2},
\end{equation}
where $c$ is the light speed in either medium (in units of the speed of light
in vacuum). The physical implications
of this condition can be found in~\cite{Lee,Brevikimpl,BrevikCJP}.
When equation (\ref{condition}) holds, we have
$\lambda_1=\lambda_2=\lambda$, and the frequency equation
(\ref{general}) is simplified considerably. It breaks down into two
equations: for the transverse-electric (TE) oscillations
\begin{equation}
\label{freq1}
\Delta ^{\text{TE}}_n(\lambda,a)\equiv \lambda a\left [
\mu_1 J'_n(\lambda a)H_n(\lambda a )-\mu_2 J_n(\lambda a)
H'_n(\lambda a)\right ]=0
\end{equation}
and for the transverse-magnetic (TM) oscillations
\begin{equation}
\label{freq2}
\Delta ^{\text{TM}}_n(\lambda,a)\equiv \lambda a\left [
\varepsilon_1 J'_n(\lambda a)H_n(\lambda a )-
\varepsilon_2 J_n(\lambda a) H'_n(\lambda a)
\right ]=0.
\end{equation}
In the general case [see Eq.~(\ref{general1})] such a decomposition occurs only
for oscillations with $n=0$. In Eqs.~(\ref{freq1}) and (\ref{freq2})
$\lambda $ is the eigenvalue of the corresponding transverse
[membrane-like] boundary value problem~\cite{LDB}
\begin{equation}
\lambda^2=\frac{\omega^2}{c^2} -k_z^2\,{.}
\end{equation}

     Classification of the solutions of Maxwell's equations without
sources in terms of the TE- and TM-modes originates in waveguide
theory~\cite{Stratton,LDB,SDMT}.  The main distinction of the propagation
of electromagnetic waves in waveguides, in contrast to the same
process in unbounded space, is that a purely transverse wave cannot
propagate in a waveguide. The wave in waveguide must necessarily
contain either longitudinal  electric or magnetic fields. The first
case is referred to as  the waves of electric type
[transverse-magnetic (TM) waves] and in the second case one is
dealing with waves of magnetic type [or transverse-electric (TE)
waves]. This classification proves to be convenient in studies of
electromagnetic oscillations in closed resonators as well.

     Replacing  the function $f_n(k_z,\omega,a)$ in Eq.~(\ref{contour})
 by the left hand sides of Eq.~(\ref{freq1}) and
(\ref{freq2}) and changing to the integration variable $\lambda $ we arrive
at the following representation for the Casimir energy
\begin{equation}
\label{contourlambda}
E=-\frac{c}{2}\int_{-\infty}^\infty \frac{dk_z}{2\pi}\sum_{n=
-\infty}^{\infty}
\frac{1}{2\pi i}\frac{1}{2}\oint_{C'}\sqrt{\lambda^2+k^2_z} \, d_\lambda
\ln\frac{\Delta_n
^{\text{TE}}(\lambda a)\Delta_n^{\text{TM}}(\lambda a)}
{\Delta_n ^{\text{TE}}(\infty)\Delta
_n^{\text{TM}}(\infty)}\,{.}
\end{equation}
     Here we have distorted the contour of integration to $C'=C'_++C'_-$.
We take $C'_+$ to consist of a straight line parallel to, and just to the right
of, the imaginary axis $(-i\infty,+i\infty)$ closed by a semicircle of an
infinitely large radius in the right half-plane.  $C'_-$ similarly is a line
parallel to, and just to the left of, the imaginary axis, closed by an
infinite semicircle in the left hand plane.  
 On both semicircles
the argument of the logarithm function in Eq.~(\ref{contourlambda})
tends to~1. As a result these parts of the contour $C'$ do not give
any contribution  to the Casimir energy~$E$. When integrating  along
the imaginary axis we chose the branch line of the function
$\varphi(\lambda)=\sqrt{\lambda^2 +k_z^2}$ to run between $-ik_z$ and $ik_z$,
where $k_z=+\sqrt{k_z^2}>0$.
In terms of $y= \text{ Im } \lambda$  we have
\begin{equation}
\varphi (iy) = \left \{
\begin{array}{lcl}
i\sqrt{y^2-k_z^2}, && y>k_z,\\
\pm\sqrt{k_z^2-y^2}, && |y|<k_z,\\
-i\sqrt{y^2-k_z^2}, && y<-k_z,
\end{array}
\right .
\end{equation}
where the sign on the middle form depends on whether we are to the right or
the left of the cut.
Thus contributions to Eq.~(\ref{contourlambda}) due to the integration
along the segment of the imaginary axis $(-i k_z,\,ik_z)$ 
cancel between $C'_+$ and $C'_-$, and Eq.~(\ref{contourlambda})
acquires the form
\begin{equation}
E=-\frac{c}{2\pi^2}\sum_{n=-\infty}^{\infty}\int_0^\infty dk_z
\int_{k_z}^\infty
\sqrt{y^2-k_z^2}\,d_y\,\ln \frac{\Delta^{\text{TE}}_n(iay)
\Delta^{\text{TM}}_n(iay)}
{\Delta_n ^{\text{TE}}(i\infty)\Delta_n ^{\text{TM}}(i\infty)}\,{.}
\end{equation}
Changing the order of integration of $k_z$ and $y$ and
taking into account the value of the integral
\begin{equation}
\int_0^y dk_z\,\sqrt{y^2-k^2_z}=\frac{\pi}{4}y^2\,{,}
\end{equation}
we obtain after substitution $a y \to y$
\begin{equation}
\label{aftersubstitution}
E=-\frac{c}{8 \pi a^2}\sum_{n=-\infty}^\infty
\int _0^\infty y^2\,d_y \, \ln
\frac{\Delta ^{\text{TE}}_n(iy)\Delta ^{\text{TM}}_n(iy)}
{\Delta ^{\text{TE}}_n(i\infty)\Delta ^{\text{TM}}_n(i\infty)}\,{.}
\end{equation}

Further  we shall need the modified Bessel functions $I_n(y)$ and
$K_n(y)$~\cite{AS}
\begin{eqnarray}
I_n(y) &=& i^{-n}J_n(iy) \\
K_n(y) &=& i^{n+1}\frac{\pi}{2}H_n(iy)
\end{eqnarray}
and their asymptotics at fixed $n$ and $y \to \infty $
\begin{equation}
I_n(y) \simeq \frac{e^y}{\sqrt{2 \pi y}},    \qquad
I'_n(y)\simeq \frac{e^y}{\sqrt{2 \pi y}},    \label{asymp1}
\end{equation}
\begin{equation}
K_n(y) \simeq \sqrt{\frac{\pi}{2y}}e^{-y}, \qquad
 K_n'(y)\simeq -\sqrt{\frac{\pi}{2y}}e^{-y}. \label{asymp2}
\end{equation}
With the help of this we derive from Eqs.~(\ref{freq1}) and
(\ref{freq2})
\begin{eqnarray}
\frac{\Delta_n ^{\text{TE}}(iy)}{\Delta^{\text{TE}}_n(i\infty)}&=&
\frac{2y}{\mu_1+\mu_2}[\mu_1 I_n'(y)K_n(y)-\mu_2I_n(y)K'_n(y)],
\nonumber\\
\frac{\Delta_n ^{\text{TM}}(iy)}{\Delta^{\text{TM}}_n(i\infty)}&=&
\frac{2y}{\varepsilon_1+\varepsilon_2}
[\varepsilon_1 I_n'(y)K_n(y)-\varepsilon_2I_n(y)K'_n(y)]\,{.}
\end{eqnarray}
Making use of all this, we can recast Eq.~(\ref{aftersubstitution}) into
the form
\begin{eqnarray}
\label{final}
E&=& \frac{c}{4\pi a^2}\sum_{n=-\infty}^\infty
\int_0^\infty y\, dy\, \ln
\biggl\{\frac{4 y^2}{\varepsilon+\varepsilon^{-1}+2}
\Bigl [\left (I_n'(y)K_n(y)\right)^2+\left (I_n(y)K_n'(y)\right )^2
 \nonumber \\
&&
-\left (\varepsilon +\varepsilon^{-1}\right )I_n(y)I_n'(y)K_n(y)K_n'(y)
\Bigr ] \biggr\}\,{.}
\end{eqnarray}
Here a new notation $\varepsilon = \varepsilon_1/\varepsilon_2$ has been
introduced, $\mu$ has been eliminated by condition (\ref{condition}),
and when going from  (\ref{aftersubstitution}) to
(\ref{final}) an integration by parts has been done, the boundary
terms being omitted. The last point will be justified further when
the removal of the divergences  is discussed. The argument  of the
logarithm in Eq.~(\ref{final}) is simplified considerably if one uses
the value of the Wronskian of the modified Bessel functions $I_n(y)$
and $K_n(y)$~\cite{AS}
\begin{equation}
\label{Wr}
I_n(y)K_n'(y) - I'_n(y)K_n(y) = -\frac{1}{y}
\end{equation}
and the identity
\begin{equation}
\label{iden}
I_n(y)K_n'(y) + I'_n(y)K_n(y) = \left(I_n(y)K_n(y)
\right )'\,{.}
\end{equation}
Finally Eq.~(\ref{final}) acquires the form
\begin{equation}
\label{finalsum}
E=\sum_{n=-\infty}^\infty E_n,
\end{equation}
where
\begin{equation}
\label{finalsimple}
E_n=\frac{c}{4\pi a^2}\int_{0}^{\infty}y \, dy \,\ln\left \{
1-\xi^2\left [y(I_n(y)K_n(y))'
\right ]^2
\right \}
\end{equation}
with $\xi =(1-\varepsilon)/(1+\varepsilon)$.  This is a simple
unregulated generalization of (4.5) of Ref.~\cite{Milton}, and the
cylindrical analog of the spherical form (7.1) of Ref.~\cite{MiltonNg}.

From the asymptotics (\ref{asymp1}) and (\ref{asymp2}) it follows
that the integral in (\ref{finalsimple}) diverges logarithmically
when $y \to \infty$. At the same time the sum over $n$ in
Eq.~(\ref{finalsum}) also diverges because at large $n$ the uniform
asymptotic expansion of the modified Bessel functions
gives~\cite{AS} (see (4.6) of \cite{Milton})
\begin{equation}
\label{infty}
 E_n \Bigr |_{n\to \infty}\simeq - \frac{c\xi^2}{16 \pi a^2}
\int _0^\infty \frac{z^5\,dz}{(1+z^2)^3}\equiv E^\infty\, {.}
\end{equation}
Here the change of variables $y=n z$ has been done. Disregarding for
the moment that the integral in Eq.~(\ref{infty}) is divergent, we employ
here the Riemann  zeta function technique~\cite{Od,El} for
attributing a finite value to the sum in (\ref{finalsum})
\begin{eqnarray}
E&=& \sum_{n=-\infty}^{\infty}(E_n-E^\infty +E^\infty)\nonumber \\
&=& \sum_{n=-\infty}^{\infty}(E_n-E^\infty) +\sum_{n=-\infty}^{\infty}
E^\infty  \nonumber \\
&=& \sum_{n=-\infty}^{\infty}\bar E_n+E^\infty\sum_{n=
-\infty}^{\infty}n^0 ,
\end{eqnarray}
where $\bar E_n$ stands for the ``renormalized''
partial Casimir energy
\begin{equation}
\label{Enr}
\bar E_n =E_n-E^\infty, \qquad n=0,\pm1, \ldots \,{.}
\end{equation}
We now have to treat the product of two divergent expressions
$E^\infty \cdot \sum_{n=-\infty}^\infty n^0$ more precisely, by presenting it
in the following form
\begin{eqnarray}
E^\infty \cdot \sum_{n=-\infty}^\infty n^0 &=&
- \frac{c \xi ^2}{16 \pi a ^2}\lim _{s \to 0^+}
\int_0^\infty\frac{z^{5-s}dz}{(1+z^2)^3}
\cdot [2 \zeta (s)+1] \nonumber \\
&=& - \frac{c \xi ^2}{16 \pi a ^2}\lim _{s \to 0^+}
  \left (\frac{1}{s} - \frac{3}{4}
\right )\cdot [2 \zeta '(s) \,s]\\
&=& - \frac{c \xi ^2}{16 \pi a ^2}\lim _{s \to 0^+}
  \left (\frac{1}{s} - \frac{3}{4}
\right )\cdot [- \ln (2\pi) s] \nonumber \\
&=& \frac{c \xi ^2}{16 \pi a ^2} \ln(2 \pi). \nonumber
\end{eqnarray}
Finally, the Casimir energy acquires the form
\begin{equation}
\label{}
E=\sum_{n=-\infty}^{\infty}\bar E_n +\frac{c\xi ^2}{16 \pi a^2}\ln (2 \pi).
\end{equation}

Now we deduce from  Eqs.~(\ref{Enr}), (\ref{finalsimple}),   and (\ref{infty})
for $\bar E_n$
\begin{equation}
\label{Enrenorm}
\bar E_n =\frac{c}{4\pi a^2}\left \{
\int_{0}^{\infty} y\, dy\, \ln
\left [1-\xi^2\sigma^2_n(y)
\right ]+\frac{\xi ^2}{4} \int_{0}^{\infty}\frac{z^5\,
dz}{(1+z^2)^3}\right \}{,}
\end{equation}
\begin{equation}
\label{Enrenorma}
\bar E_{-n}=\bar E_n, \qquad n=0,1,2,\ldots \,{,}
\end{equation}
where $\sigma _n(y) =y (I_n(y)K_n(y))'$.  Since both integrals
in Eq.~(\ref{Enrenorm}) rdiverge,
the finite sum is to interpreted in a precise manner as specified below.

     The removal of the divergences by making use of the
$\zeta$~function justifies dropping the boundary terms in the
integration by parts when we went from Eq.~(\ref{aftersubstitution})
to (\ref{final}). To see this one can at first remove the divergences
in Eq.~(\ref{aftersubstitution}) by employing the Riemann $\zeta
$~function as described above. After that the integration by parts in
the formula for the renormalized partial Casimir energy $\bar E_n$
can be done, with  boundary terms now vanishing rigorously. Proceeding
in that way one again arrives at  Eqs.~(\ref{Enrenorm}) and
(\ref{Enrenorma}).

	In order for  Eq.~(\ref{Enrenorm}) to be cast in a form
suitable for numerical evaluations both terms there should be placed
under s single integral sign. To this end, for $n\ne 0$ the
change of variables $z=y/n$  has to be done in the second term.
This change is inverse to the one which has led to the asymptotic form
(\ref{infty}).  This yields
\begin{equation}
\label{Enfinal}
\bar E_n =\frac{c}{4\pi a^2}\int_{0}^{\infty} y\, dy\,\left \{ \ln
\left [1-\xi^2\sigma^2_n(y)
\right ]+\frac{\xi ^2}{4} \frac{y^4}{(n^2+y^2)^3}\right \},
\quad n=1,2,\ldots {.}
\end{equation}
For $n=0$ Eq.~(\ref{Enrenorm}) is rewritten as
\begin{equation}
\label{E0final}
\bar E_0 =\frac{c}{4\pi a^2}\int_{0}^{\infty} y\, dy\,\left \{ \ln
\left [1-\xi^2\sigma^2_0(y)
\right ]+\frac{\xi ^2}{4} \frac{y^4}{(1+y^2)^3}\right \}\,{.}
\end{equation}
The integrals in these formul\ae\ converge because for $y\to 0$ and $n\ne 0$
we have~\cite{AS}
\begin{equation}
I_n(y)K_n(y) \to \frac{1}{2n}\,{.}
\end{equation}
In this limit
\begin{equation}
\sigma ^2_0(y)\to 1\,{.}
\end{equation}
On the other hand, for large $y$
\begin{equation}
\sigma ^2_n(y)\to \frac{1}{4y^2}\,{.}
\end{equation}

By making use of uniform  asymptotics of the Bessel
functions~\cite{AS} we deduce from~(\ref{Enfinal})
\begin{equation}
\label{Enasymp}
\bar E_n\Bigl |_{n\to \infty}\sim \frac{c \xi^2}{4\pi a^2}
\left[\frac{10-3 \xi ^2}{960n^2}-\frac{28224-7344\xi^2+720\xi^4}{15482880n^4}
+O\left(\frac{1}{n^6}\right)\right] \,{.}
\end{equation}
Thus the Casimir energy
\begin{equation}
\label{Efinal}
E=\bar E_0+2\sum_{n=1}^{\infty}\bar E_n+\frac{c\xi ^2}{16 \pi a^2}
\ln (2 \pi)
\end{equation}
with $\bar E_0$   and $\bar E_n$, $n\ge 1$, given in (\ref{E0final}) and
(\ref{Enfinal}), respectively, is finite~\cite{endnote2}. One can
advance further only by considering special cases and applying numerical
calculations.
\section{Dilute compact cylinder and perfectly conducting cylindrical
shell}
     We begin by addressing the case when $\xi ^2\ll 1$.
Because we are assuming the condition
$\varepsilon_1\mu_1=\varepsilon_2\mu_1= c^{-2}$ and $\xi^2\ll 1$, this is
not the same situation as a dilute compact cylinder with
$|\varepsilon_1-\varepsilon_2|\ll 1$ and $\mu_1=\mu_2=1$.  In this case
\begin{equation}
\xi ^2=\left( \frac{\varepsilon_1-\varepsilon_2}{\varepsilon_1
+\varepsilon_2}
\right )^2=\frac{(\varepsilon_1-\varepsilon_2)^2}{4\varepsilon^2},
\end{equation}
where $\varepsilon= (\varepsilon_1+\varepsilon_2)/2$.
Retaining in Eq.~(\ref{E0final}) only the terms proportional to $\xi
^2$ we  obtain
\begin{eqnarray}
\label{E0dilute}
\bar E_0&\simeq&\frac{c \xi^2}{4\pi a^2} \int_{0}^{\infty}y\,dy\left [
\frac{y^4}{4(1+y^2)^3}-\sigma _0^2(y)
\right ]                             \nonumber \\
&\simeq & \frac{c\xi ^2}{4\pi a^2}\,(-0.490878)\,{.}
\end{eqnarray}
To estimate $\bar E_n$, $n >0$, we can use the leading asymptotic behavior
 (\ref{Enasymp})
\begin{equation}
\label{Endilute}
\bar E_n\sim \frac{c\xi^2}{4\pi a^2}\left(\frac{1}{96 n^2}
-\frac{7}{3840n^4}\right)\,{.}
\end{equation}
To a precision of $10^{-6}$, we evaluate Eq.~(\ref{Efinal})
by substituting in the value of $\bar E_0$, (\ref{E0dilute}),
integrating $\bar E_n$ numerically for $n=1,\dots 5$, and
asymptotically using Eq.~(\ref{Enasymp}) for $n\ge6$, with the
result\footnote{The cancellations here are very severe.
If the asymptotic approximation were used for all $n$, a positive result
would be found, $E\sim (c\xi^2/4\pi a^2)(-0.00108)$.
Unlike for the spherical case, doing the integral exactly is essential.}
\begin{eqnarray}
E&\simeq &
\frac{c \xi ^2}{4\pi a^2}\left (-0.490878 + 2\sum_{n=1}^5\bar E_n
+\frac{1}{48}\sum_{n=6}^\infty \frac{1}{n^2}
-\frac{7}{1920}\sum_{n=6}^\infty\frac{1}{n^4}+ \frac{1}{4} \ln (2 \pi)
\right )\nonumber \\
&=&
\frac{c \xi ^2}{4\pi a^2}\left (-0.490878 +0.027638 +
0.003778-0.000007+0.459469 \right )\\
&=& \frac{c \xi ^2}{4\pi a^2}(0.000000) \,{.} \nonumber
\end{eqnarray}
Thus the Casimir energy of a cylinder possessing the same speed of light
inside and outside  proves to be zero!  This is to be contrasted with
the positive Casimir energy found for a dilute ball with the same
property~\cite{BrevikK,MiltonNg,Ball},
\begin{equation}
E_{\rm ball}\approx {3\over64a}\xi^2=0.046875{\xi^2\over a}.
\end{equation}
It is further remarkable that the same zero result is found for a
dilute {\it dielectric\/} cylinder, that is, one with $\mu=1$ everywhere
and $\varepsilon>1$ inside the cylinder, and $\varepsilon=1$ outside, a result
which may be most easily confirmed by summing the intermolecular
van der Waals energies.
That calculation is given in Appendix B.  However, zero is not the
universal value of the Casimir energy for cylinders, as we now remind the
reader.

Of particular interest is the case when $\xi^2=1$. With  $c=1$ in
our formulas it corresponds to infinitely  thin and perfectly
conducting cylindrical shell (see Appendix C).
Setting  $\xi =1$ and $c=1$ in Eq.~(\ref{E0final}) we obtain
by numerical integration\footnote{In the notation of Ref.~\cite{Milton}
this is $-{1\over8\pi a^2}(S+R_0+{1\over2}\ln 2\pi)$---see Eqs.~(5.5) and (5.6)
of Ref.~\cite{Milton}.  The $\ln2\pi$ term is cancelled by that in
Eq.~(\ref{Efinal}) here.}
\begin{equation}
\label{E0shell}
\bar E_0= \frac{1}{4\pi a^2}(-0.6517)=-0.05186\,\frac{1}{a^2}\,{.}
\end{equation}

The sum $2\sum_{n=1}^{\infty}\bar E_n$  in Eq.~(\ref{Efinal}) can be
found  by making use of the two leading terms in the uniform
asymptotic expansion
\begin{eqnarray}
\label{sumshell}
2\sum_{n=1}^{\infty}\bar E_n&\simeq &\frac{1}{4\pi a^2} \left (
\frac{7}{480}
\sum_{n=1}^{\infty}\frac{1}{n^2} - \frac{5}{1792} \sum _{n=1}^{\infty}
\frac{1}{n^4}\right )=
\frac{1}{4\pi a^2}\left (
\frac{7}{480}\frac{\pi ^2}{6}-\frac{5}{1792}\frac{\pi ^4}{90} \right )\nonumber
\\
&=&\frac{1}{4\pi a^2}0.0210 = 0.0018\,\frac{1}{a^2}\,{.}
\end{eqnarray}
With higher accuracy (up to $10^{-5}$) this sum has been
calculated in~\cite{Milton} by
integration of~(\ref{Enfinal})\footnote{This is exactly the same as $-{1\over
8\pi a^2}R$ given by Eq.~(5.11) of Ref.~\cite{Milton}.}
\begin{equation}
\label{Miltoncal}
2\sum_{n=1}^{\infty}\bar E_n\simeq\frac{1}{4\pi a^2}\,\frac{1}{2}\,0.0437=
\frac{1}{4\pi a^2}\,0.0218= 0.00174
\,\frac{1}{a^2}\,{.}
\end{equation}
Substituting  Eqs.~(\ref{E0shell}) and (\ref{Miltoncal}) into
(\ref{Efinal}) we obtain for the Casimir energy of a perfectly
conducting cylindrical shell
\begin{equation}
\label{Eshell}
E_{\text{shell}}=\frac{1}{4\pi a^2}\,(-0.1704)=-0.01356\,
\frac{1}{a^2}\,{.}
\end{equation}
This is exactly the result first obtained by DeRaad
and Milton~\cite{Milton,Raad}.
It is worth noting here that unlike in that approach the use of the
$\zeta$ function techniques
enables us to dispense with introduction of the high-frequency
cutoff function, although the latter is undoubtedly more physical.

While this paper was being completed the authors became
 aware of Ref.~\cite{GR}
where the vacuum energy of an infinite perfectly conducting cylindrical
surface has also been rederived, to much higher accuracy, but using
a rather more elaborate method.

\section{conclusion}
     When calculating the Casimir energy there appear to be many
arbitrary methods of controlling and removing divergences.
It is therefore reassuring that unique results emerge whatever
regularization scheme is adopted.  This is somewhat less trivial
for this cylindrical case than for the case of a sphere, because of the
subtleties associated with even space dimensions \cite{MiltonPR}.
However, this universality seems to be characteristic of Casimir calculations
even in cases where finiteness is not achieved, as for a
dielectric ball without the condition (\ref{condition}) imposed
\cite{bmm,barton}.  Thus our understanding of the Casimir effect seems to be
improving.  However, this comforting conclusion must be tempered by the
surprising new result that for both dilute dielectric-diamagnetic (satisfying
$\varepsilon\mu=\mbox{const.}$) cylinders, and for dilute dielectric cylinders,
the Casimir energy vanishes.  We have as yet no theoretical understanding
of these zeroes.

\acknowledgments
We thank M. Bordag for organizing the Fourth Workshop on Quantum Field
Theory Under the Influence of
External Conditions, which enabled the collaboration to
take place.  We are particularly grateful to A. Romeo for pointing out
that the dielectric cylinder had vanishing Casimir
energy, and thereby triggering
the detection of a transcription error in the calculation of the
energy for a dilute dielectric-diamagnetic cylinder.
This work was accomplished with partial financial support of the Russian
Foundation of Fundamental Research (Grant No.~97-01-00745) and of the
U.S. Department of Energy (Grant No.~DE-FG-03-98ER41066).

\appendix
\section{Relation of Method to Green's Function Approach}

Here we sketch the relation of the formula (\ref{contour}) to the
Green's function formalism.  We will confine our remarks to the
case of a massless scalar field, as the generalization to, say, electromagnetism
is rather immediate.  We take the scalar Green's function to satisfy
the differential equation
\begin{equation}
\left({\partial^2\over\partial t^2}-\nabla^2\right)G(x,x')=-\delta(x-x'),
\label{gfneqn}
\end{equation}
subject to appropriate boundary conditions.  The stress tensor (use of
the conformal stress tensor has the same effect) is
\begin{equation}
t^{\mu\nu}=\partial^\mu\phi\partial^\nu\phi-{1\over2}g^{\mu\nu}\partial_\lambda
\phi\partial^\lambda\phi,
\end{equation}
so, since the Green's function is given as a vacuum expectation value of a
time-ordered product of fields,
\begin{equation}
G(x,x')=-i\langle T\phi(x)\phi(x')\rangle,
\end{equation}
the energy density is
\begin{eqnarray}
u=\langle t^{00}\rangle=-{i\over2}[\partial^0\partial^{0\prime}+\bbox{\nabla}
\cdot\bbox{\nabla}']G(x,x')\bigg|_{x=x'}.
\end{eqnarray}
To compute the total energy, we integrate $u$ over all space; then we can
integrate by parts on one of the gradient terms, and use the differential
equation (\ref{gfneqn}), omitting the delta function because the limit of
point coincidence is understood:
\begin{equation}
\bbox{\nabla}\cdot\bbox{\nabla}'\to-\nabla^2\to-{\partial^2\over\partial t^2}.
\end{equation}
(The net effect is that the Lagrangian term in $t^{\mu\nu}$ does not
contribute.)  In terms of the Fourier transform of the Green's function
\begin{equation}
G(x,x')=\int{d\omega\over2\pi}e^{-i\omega(t-t')}{\cal G}_\omega({\bf r,r'}),
\end{equation}
the Casimir energy is
\begin{equation}
E=-i\int{d\omega\over2\pi}e^{-i\omega(t-t')}\int(d{\bf r})\,
\omega^2{\cal G}_\omega({\bf r,r'})\bigg|_{x=x'}.
\end{equation}

Now introduce eigenfunctions of the differential operator $\nabla^2$
subject to the same boundary conditions as ${\cal G}_\omega$:
\begin{eqnarray}
\nabla^2\psi({\bf r})&=&-k^2_p\psi({\bf r}),\\
\sum_p\psi_p({\bf r})\psi_p^*({\bf r}')&=&\delta({\bf r-r'}),\\
\int(d{\bf r})\psi_p^*({\bf r})\psi_{p'}({\bf r})&=&\delta_{pp'}.
\end{eqnarray}
Then the Green's function has an eigenfunction expansion
\begin{equation}
{\cal G}_\omega({\bf r,r'})=\sum_p{\psi_p({\bf r})\psi_p^*({\bf r'})\over
\omega^2-k_p^2}.
\end{equation}
Carrying out the volume integration,
\begin{equation}
\int(d{\bf r}){\cal G}_\omega({\bf r,r})=\sum_p{1\over\omega^2-k_p^2},
\end{equation}
we find that the energy can be written in the form
\begin{equation}
E=-i\int_{-\infty}^\infty {d\omega\over2\pi}e^{-i\omega \tau}{\omega\over2}
\sum_p\left({1\over\omega-k_p}+{1\over\omega+k_p}\right).
\end{equation}
Here we have retained a time splitting, $\tau=t-t'\to 0$, which is a
technique to regulate the divergent expression. What does this integral mean?
Since the energy must be real, when $\tau$ is set equal to zero, 
the $-i$ is to be interpreted as an instruction to pick out the
negative imaginary part of the integral.  That means that the contour
of integration $C$ must encircle all the poles on the real axis, the positive
poles by a contour closed in the counterclockwise sense, and the
negative poles by a contour closed in the clockwise sense: (see Ref.~\cite{km})
\begin{equation}
E=-{i\over8\pi}\int_{C}\omega \, d_\omega\ln\prod_p(\omega-k_p)(\omega+k_p).
\end{equation}
(We may verify the sign by noting that Eq.~(\ref{definition}) 
is formally reproduced.) This is the content of (\ref{contour}).

\section{Van der Waals energy of a dielectric cylinder}

It is now established that for tenuous media the Casimir effect and
the sum of molecular van der Waals forces are identical \cite{bmm}.
Here we calculate the latter for a dilute solid cylinder, with
dielectric constant $\varepsilon\ne1$ in the interior, $\varepsilon=1$ in
the exterior, and $\mu=1$ everywhere.  We follow the procedure given in
Ref.~\cite{MiltonNg2}.  The van der Waals energy for this sphere is
\begin{equation}
E_{\rm vdW}=-{1\over2}BN^2\int d^Dr\,d^Dr'[|{\bf r_\perp-r_\perp'}|^2
+r^2+r^{\prime2}-2rr'\cos\theta]^{-\gamma/2},
\end{equation}
where $B=(23/4\pi)\alpha^2$, $\alpha=(\varepsilon-1)/4\pi N$ being the
molecular polarizability, and $N$ being the number density of molecules.
We have regulated the integral by dimensional continuation, $D$ being
the number of spatial dimensions, and $\gamma$ being the (inverse) power
of the Casimir-Polder potential.  The following calculation is valid
providing $D>\gamma$; the final result will be obtained by violating
this condition, by setting $D=3$ and $\gamma=7$.

We assume translational invariance in the $D-2$ transverse directions,
so the transverse integral is easy: ($L$ is the length of the cylinder
and $b^2=r^2+r^{\prime 2}-2rr'\cos\theta$)
\begin{eqnarray}
\int_{-\infty}^\infty d^{D-2}r_\perp \,d^{D-2}r_\perp'\,[|{\bf r_\perp-r_\perp'}
|^2+b^2]^{-\gamma/2}&=&L^{D-2}\int_{-\infty}^\infty d^{D-2}r_\perp\,
[r_\perp^2+b^2]^{-\gamma/2}\nonumber\\
&=&{L^{D-2}\over\Gamma(\gamma/2)}\int_{-\infty}^\infty d^{D-2}r_\perp
\int_0^\infty {dt\over t}t^{\gamma/2}e^{-t(r^2_\perp+b^2)}\nonumber\\
&=&{L^{D-2}\over\Gamma(\gamma/2)}\int_0^\infty dt\, t^{\gamma/2-1}e^{-tb^2}
\left[\int_{-\infty}^\infty dx \,e^{-tx^2}\right]^{D-2}\nonumber\\
&=&(L\sqrt{\pi})^{D-2}(b^2)^{D/2-\gamma/2-1}{\Gamma(\gamma/2-D/2+1)\over
\Gamma(\gamma/2)}.
\end{eqnarray}
The remaining integral over $r$, $r'$, $\theta$, $\theta'$ is just that
given in Ref.~\cite{MiltonNg2}.  In Eq.~(3.25) there, we merely set $D=2$
and $\gamma=\gamma-D+2$.  The result is
\begin{equation}
E_{\rm vdW}=-BN^2{L^{D-2}\over a^{\gamma-D-2}}{2^{D-\gamma}\pi^{D/2+1/2}
\Gamma(\gamma/2-D/2+1)\Gamma(D/2-\gamma/2+1/2)\over
\Gamma(\gamma/2)\Gamma(D/2-\gamma/2+2)(D-\gamma)}.
\end{equation}
This is exactly the result found by Romeo \cite{romeo}.  Now when we
set $D=3$ and $\gamma=7$ everything is finite except for the second
gamma function in the denominator, which has a simple pole, and thus
the Casimir energy vanishes in this case.

\section{Infinitely thin perfectly conducting cylindrical shell}
We show here that the Casimir energy for an infinitely thin 
perfectly conducting cylindrical shell is given by
Eqs.~(\ref{finalsum}) and (\ref{finalsimple}) with $\xi ^2=1$. In
this case, as for a perfectly conducting  spherical shell, the
frequencies of electromagnetic oscillations inside and outside the
shell turn out to be different~\cite{Stratton}. The frequencies
of the TE-modes are determined by\footnote{The
exterior modes can be considered only formally, since the Hankel functions
have only a finite number of {\it complex\/} zeroes.  See Ref.~\cite{hagen}.
Nevertheless, this formal procedure yields the correct result.  The mode sum
breaks down in this case, because the singularity structure is not that
assumed in (\ref{contour}), but the Green's function method retains meaning.}
\begin{eqnarray}
J'_n(\lambda a)&=&0, \quad r<a, \label{A1} \\
 H'_n(\lambda
a)&=& 0, \quad r >a\,{.}       \label{A2}
\end{eqnarray}
For the TM-modes we have
\begin{eqnarray}
J_n(\lambda a)&=&0, \quad r<a, \label{A3} \\
H_n(\lambda a)&=& 0, \quad r >a\,{.}     \label{A4}
\end{eqnarray}
In these equations
\begin{equation}
\lambda^2=\omega ^2-k^2_z, \qquad n=0,\pm 1,\pm 2,\ldots \,{.}
\end{equation}
Substituting  $\Delta^{\text{TE}}_n$
and $\Delta^{\text{TE}}_n$ in Eq.~(\ref{aftersubstitution})
by the new equations (\ref{A1})--(\ref{A4}) we
obtain [cf.\ with Eq.~(\ref{final})]
\begin{equation}
E=\frac{1}{4\pi a^2}\sum_{n=-\infty}^{\infty}\int_0^\infty y\,dy\,\ln\left\{
-4y^2I_n(y)I_n'(y)K_n(y)K_n'(y)\right\}\,.
\end{equation}
To rearrange the argument  of the logarithmic function in this
formula  we  again use equalities (\ref{Wr}) and (\ref{iden}). This
gives Eqs.~(\ref{finalsum}) and (\ref{finalsimple}) with $\xi^2=1$
and~$c=1$:
\begin{equation}
E=\frac{1}{4\pi a^2}\sum_{n=-\infty}^{\infty}\int_0^\infty y\, dy \,
 \ln \left \{
1-\left [  y(I_n(y)K_n(y))'\right ]^2
\right\}\,,
\end{equation}
which is, of course, the unregulated version of the result derived rigorously
in Ref.~\cite{Milton}.


%
%

%
%


\begin{references}
\bibitem{BD} R.~Balian and B.~Duplantier, Ann.\ Phys.\ (N.Y.) {\bf
112}, 165 (1978).
\bibitem{Milton} L.~L.~DeRaad, Jr.\ and K.~Milton, Ann.\ Phys.\ (N.Y.)
{\bf 136}, 229 (1981).
\bibitem{Sphere} V.~V.~Nesterenko and I.~G.~Pirozhenko,
Phys.\ Rev.~D {\bf 57}, 1284 (1998).
\bibitem{Ball}  I.~Brevik,  V.~V.~Nesterenko, and I.~G.~Pirozhenko,
 {\it Direct
mode summation for the Casimir energy of a solid ball},
Report No.\ JINR
E2-97-307, Dubna, 1997, hep-th/9710101; to be published in J.~Phys. A: Math.\
Gen.
\bibitem{zeta}  V.~V.~Nesterenko and I.~G.~Pirozhenko,
 J.~Math.\ Phys.\ {\bf 38}, 6265 (1997).
\bibitem{hagen} M. E. Bowers and C. R. Hagen, ``Casimir Energy of a
Spherical Shell,' preprint UR-1533.
\bibitem{BN} I.~Brevik and G.~H.~Nyland, Ann.\ Phys.\ (N.Y.) {\bf
230}, 321 (1994).
\bibitem{Raad} L.~L.~DeRaad, Jr., Fortschr.\ Phys.\ {\bf 33}, 117
(1985).
\bibitem{MRS} K.~A~Milton, L.~L.~DeRaad Jr., and
J.~Schwinger, Ann.  Phys. (N.Y.) {\bf 115}, 338 (1978).
\bibitem{Miltonb} K.~A~Milton, Ann.\ Phys.\ (N.Y.) {\bf 127}, 49
(1980).
\bibitem{ML} K.~A.~Milton, in {\it Proceedings of the Third
 Workshop
on Quantum Field Theory Under the Influence of External
Conditions, Leipzig, 1995}, edited by M.~Bordag, (Teubner, Stuttgart,
1996),
p.~13.
\bibitem{MiltonNg} K.~A.~Milton and Y.~J.~Ng, Phys. Rev. E {\bf 55},
4207 (1997).
\bibitem{MiltonNg2} K.~A.~Milton and Y.~J.~Ng,
Phys. Rev. E {\bf 57}, 5504 (1998).
\bibitem{BrevikK} I.~Brevik and H.~Kolbenstvedt, Ann. Phys. (N.Y.)
{\bf 143},
179 (1982); {\bf 149}, 237 (1983).
\bibitem{Brevikcor} I.~Brevik, J.~Phys.~A: Math.\ Gen.\ {\bf 20},
5189 (1987).
\bibitem{Stratton} J.~A.~Stratton, {\it Electromagnetic Theory}
(McGraw-Hill, New York, 1941).
\bibitem{endnote} In Ref.~\cite[Section 9.15]{Stratton}
Eq.~(\ref{general}) is considered in
connection with the propagation of electromagnetic waves of
frequency $\omega$ along a solid cylinder. As a consequence it is
treated there as an equation determining the wave number $k_z$ at
a given $\omega$.
\bibitem{Lee} T.~D.~Lee, {\it Particle Physics and Introduction to
Field Theory} (Harwood, New York, 1981).
\bibitem{Brevikimpl} I.~Brevik,  J.~Phys.\ A: Math.\ Gen.\ {\bf 15},
L369 (1982); Can.\ J.\ Phys.\ {\bf 61}, 493 (1983).
\bibitem{BrevikCJP} I.~Brevik and H.~Kolbenstvedt, Can.\ J.\ Phys.\
{\bf 62},
805 (1984); {\bf 63}, 1409 (1985); Phys.\ Rev.\ D {\bf 25}, 1731
 (1982).
\bibitem{LDB} Louis De Broglie, {\it Problems de propagations
guidees des ondes
electromagnetiques} (Gauthiers-Villars, Paris, 1941).
\bibitem{SDMT} J. Schwinger, L. L. DeRaad, Jr., K. A. Milton, and W.-y. Tsai,
{\it Classical Electrodynamics\/} (Perseus, Reading, Massachusetts, 1998).
\bibitem{AS} M.~Abramowitz and I.~A.~Stegun, {\it Handbook of
Mathematical Functions} (National Bureau of Standards,
Washington, D.\ C., 1964; reprinted by Dover, New York, 1972).
\bibitem{Od} E.~Elizalde, S.~D.~Odintsov, A. Romeo, A.~A.\
 Bytsenko, and S.~Zerlini, {\it Zeta regularization techniques
 with applications} (World Scientific, Singapore, 1994).
 \bibitem{El}
E.~Elizalde, {\it Ten Physical Applications of Spectral Zeta
Functions} (Springer, Berlin, 1995).
\bibitem{endnote2} Applying the Green's function techniques
this problem
has been considered in~\cite{BN}.
\bibitem{GR} P.~Gosdzinsky and A.~Romeo, {\it Energy of the vacuum with a
perfectly conducting and infinite cylindrical surface}, hep-th/9809199.
\bibitem{MiltonPR} C.~M.~Bender and K.~A.~Milton, Phys.\ Rev.\ D {\bf
50}, 6547 (1994); K. A. Milton, {\it ibid.} {\bf 55}, 4940 (1997).
\bibitem{bmm} I. Brevik, V. N. Marachevsky, and K. A. Milton,
``Identity of the van der Waals and the Casimir Effect and the
Irrelevance of these Phenomenon to Sonoluminescence,'' OKHEP-98-08.
\bibitem{barton} G. Barton, ``Perturbative Check on the Casimir Energy
of a Nondispersive Dielectric Ball.''
\bibitem{km} R. Kantowski and K. A. Milton, {\it Phys. Rev.} D {\bf 35}, 549
(1987).
\bibitem{romeo} A. Romeo, private communication.

\end{references}
\end{document}